\documentclass[10pt]{article}
\newtheorem{ex}{Example}
\title{Extending Bell inequalities to more parties}
\author{Yu-Chun Wu$^{1,2,5}$, Piotr Badziag$^{3,1,5}$, Marcin Wie\'sniak$^{1,4,5}$ and Marek \.Zukowski$^{1,3,5}$\\
\small $^1$Institute of Theoretical Physics and Astrophysics, University of Gda\'nsk, PL-80-952 Gda\'nsk\\
\small $^2$Key Laboratory of Quantum Information, \\ \small University of Science and Technology of China, 230026 Hefei, China\\
\small $^3$Alba Nova Fysikum, University of Stockholm, S106 91, Sweden\\
\small $^4$Department of Physics, National University of Singapore, 2 Science Drive 3, Singapore 117542\\
\small $^5$National Quantum Information Centre of Gdansk, ul. W. Andersa 27, PL\~81-824 Sopot, Poland}
\date{}

\begin{document}

\maketitle

\begin{abstract}

We describe a method of extending Bell inequalities from $n$ to $n+1$ parties and formulate sufficient conditions for our method to produce tight inequalities from tight inequalities. The method is non trivial in the sense that the inequalities produced by it, when applied to entangled quantum states may be violated stronger than the original inequalities. In other words, the method is capable of generating inequalities which are more powerfull indicators of non-classical correlations than the original inequalities.

\end{abstract}

\section{Introduction}
Tight Bell inequalities mark the limitations of local realism. Thus,
to fully understand the boundary between the classical and the
quantum, one would like to generate a complete set of these inequalities. Moreover, apart from a purely cognitive, there is also a utilitarian aspect to this effort: each distributed quantum state, which
violates a Bell inequality reduces complexity of some quantum-communication tasks \cite{BZPZ04}. In other words, each quantum state $\rho$, which violates a Bell inequality is potentially a more useful resource for modern information technology than any state containing classical correlations only.

A fruitful approach to the search for new Bell inequalities relies on the fact that locally realistic correlations can be associated with polytopes defined by their vertices. In this picture tight Bell inequalities
represent the facets of the polytopes. To find the facets from the vertices may seem elementary, nevertheless it is a computationally NP-hard problem \cite{pit2}. Hence, to find the full set of  Bell
inequalities is a difficult task. Nevertheless, in some cases the
task can be manageable. There are algorithms, which find facets of
polytopes  from their vertices. These algorithms can be used to find
such tight Bell inequalities, which correspond to low-dimensional
polytopes \cite{tsir}. There are even relative software
packages to compute and classify Bell inequalities in low dimensions
\cite{pit,sliwa}. Some of them use the triangle elimination \cite{cutpoly,cutpoly2} and lifting \cite{sp} in convex polytopes to obtain new Bell inequalities by suitable
deformations of the old ones. 

Moreover, the search for many Bell inequalities (all for up to 3 experimental settings per observer)
can be reduced to the search for some well defined sign functions
\cite{zuk0,zuk,wbz}. In case of the inequalities involving only two measurement settings per observer, this search can be further simplified, so that in this case one can construct a complete set of
Bell inequalities with arbitrary many observers \cite{ww, ZB01}. Nevertheless, generation of multi-observer inequalities with more than three measurement settings per observer still remains a difficult task. For some methods of obtaining such inequalities see \cite{wz}.

In this paper we propose a simple method to generate multi-observer Bell inequalities from the inequalities involving fewer observers. Our method transforms tight inequalities into tight inequalities. Moreover, the extensions are non trivial in the sense that they may increase the maximum degree of violation of the inequality by the measurements on entangled quantum states.

\section{Correlation polytope}

Our considerations are restricted to inequalities describing binary local observables (possible
outcomes $\pm 1$). The parties $A$(lice), $B$(ob), $C$(harlie), $\ldots$ are then allowed to measure $M, \ N, \ K, \ \ldots$  local observables respectively.

By $A_1,A_2,\ldots,A_M$, $B_1,B_2,\ldots,B_N$, etc. we
denote the observables used by Alice, Bob, etc. respectively and by $a_i$, $b_j$, etc.
we denote the measurement outcomes corresponding to the measurement settings
$A_i$, $B_j$, etc respectively.

With this notation, all possible realistic models describing 2-party correlations must have their predictions inside a $MN$-dimensional polytope (the correlation polytope), with the vertices
$$
\vec{a}\otimes\vec{b}=(a_1,a_2,\ldots,a_M)\otimes
(b_1,b_2,\ldots,b_N),
$$
where for all $i$ and $j$ the components $a_i,\, b_j$ are $\pm 1$. In what follows, vectors with components $\pm 1$  referred to as admissible vectors.

Because the correlation polytope has inversion symmetry about the
origin, none of its faces may cross the origin. We may thus
assume that every equation, which describes a hyperplane containing a face of the polytope is of the form
\begin{equation}\label{hyperequ}
I(\vec{a},\vec{b})=1,
\end{equation}
where $I(\vec{a},\vec{b})$ is a linear function of $\vec{a}\otimes\vec{b}$, i.e., $I(\vec{a},\vec{b})= \sum_{ij}{F_{ij} a_i b_j}$. The hyperplane contains a facet of the polytope if and only if there are $M N$ linearly independent vertices $\{\vec{a}_s\otimes\vec{b}_s\}$ satisfying $I(\vec{a}_s,\vec{b}_s)=1$ and for the remaining vertices $I(\vec{a},\vec{b})\leq 1$ \cite{math}. In general, it is possible that the number of vertices saturating equation (\ref{hyperequ}) is larger than the dimension of the underlying space (in our case $MN$). For example, a facet of a cube in $\mathcal{R}^3$ contains four vertices. The number of linearly independent vertices among them is however 3, which is equal to the dimension of the space. Hence, existence of $M N$ linearly independent vertices,
satisfying equation (\ref{hyperequ}), implies that the equation
describes a facet. In other words, it immediately leads to an optimal (tight) Bell inequality.

\section{The method}

The idea behind the method is rather simple. It can be illustrated by considering $3$-particle extensions of the famous CHSH inequality (c.f. \cite{ww}). The CHSH inequality limits local realistic
correlations between measurement results of two observers $A$ and
$B$. Each of the observers measures a dichotomic observable
(possible outcomes $\pm 1$) in one of two measurement settings:
$A_1,\,A_2$ and $B_1,\,B_2$ respectively. The associated
measurement outcomes are then naturally denoted by $a_1,\,a_2$ and $b_1,\,b_2$ respectively.  In this notation, the CHSH inequality reads
$$
I(\vec{a}, \vec{b})=\frac{1}{2}\left<(a_1 + a_2)b_1 + (a_1 - a_2)b_2\right> \leq 1.
$$

One can easily convince oneself that this inequality is tight, i.e., its saturation describes a facet of the corresponding correlation polytope. Indeed, out
of the eight vertices (possible sets of locally realistic products) $(a_1,\ a_2) \otimes (b_1,\ b_2) = (1,\ \pm1) \otimes (\pm 1,\ \pm1)$ of its $4$-dimensional polytope, four vertices $(1,1) \otimes (1,1)$, $(1,1) \otimes (1,-1)$, $(1,-1) \otimes (1,1)$ and $(1,-1) \otimes (-1,1)$ saturate the inequality and it is easy to see that
these four vertices represent linearly independent vectors. This is enough for the four vertices to generate a facet of the polytope.

Clearly, if $b_1$ and $b_2$ are replaced by sign functions of two
variables, $b_1'(\vec{b},\vec{c})$ and $b_2'(\vec{b},\vec{c})$, then
one will obtain a valid Bell inequality too. The new inequality will
limit local realistic correlations between measurements of three
observers. In general, the new inequality will not be tight. The
challenge then is to formulate guidelines for the generation of
tight inequalities in this way.

As the first step toward this goal, let us consider the following simple transformation:
\begin{eqnarray*}
b_1^{\prime} &=& b_1c_1,\\
b_2^{\prime} &=& b_2c_1,\\
\end{eqnarray*}
The transformation can be put in the form $\vec{b}^{\prime}=U(\vec{c}) \vec{b}$ with a diagonal transformation matrix
$$
U(\vec{c})=\left(\begin{array}{cc}
                  c_1 & 0 \\
                  0 & c_1 \\
\end{array}\right),
$$

Matrix $U$ has two easily noticeable properties:
\begin{description}
    \item[(a)]  For every admissible vector $\vec{c}$, $U(\vec{c})$ transforms admissible vectors into new admissible vectors.
    \item[(b)]  For every admissible vector $\vec{c}$, $U(\vec{c})$ is non-singular.
 \end{description}

One can easily convince oneself that when these two conditions for $U(\vec{c})$ are satisfied then the resulting $3$-observer Bell inequality is tight (the same applies to the case when $U(\vec{c})$ is used to produce a $p+1$-observer inequality from a $p$-observer inequality for arbitrary $p$). On the other hand, none of the two conditions is necessary to produce a $p+1$-observer tight inequality from a given tight $p$-observer inequality. Thus, we will be able to relax condition (b). We will, however, keep condition (a).

Condition (a) is convenient since its fulfillment is necessary for $U(\vec{c})$ to produce valid $3$-party Bell inequalities from all $2$-party inequalities (the argument easily generalizes to more parties as well). Moreover the condition excludes a large number of extensions, which clearly produce non-optimal inequalities from optimal (tight) ones.

Indeed, assume that for some admissible vector $\vec{b}$, transformation $U(\vec{c})$ produces $\vec{b}'$ with the first component $b'_1 > 1$ and consider the extension of a trivial $2$-party inequality $a_1 b_1 \leq 1$. The extension reads $a_1 b'_1(\vec{b},\vec{c}) \leq 1$. Now, due to the assumption, there are $3$-party vertices, for which $a_1=1$ and $b'_1 > 1$, i.e., the inequality is not satisfied. A similar argument shows that when $U(\vec{c})$ produces a vector $\vec{b}'$ with $b'_1 < 1$ then the extension of a trivial inequality $a_1 b_1 \leq 1$ is not tight.

Condition (a) puts a very strong limitation on the elements of $U(\vec{c})$. To begin with, one can easily convince oneself that the condition requires that for every admissible vector $\vec{c}$, each row of $U(\vec{c})$ contains one element equal to $\pm 1$ and the remaining elements of the row are zero. 
This, in turn, implies that each row may have either one or two elements, which are not identically zero. In the former case, the non-zero element is equal to one of the components of vector $\vec{c}$, say $c_k$ multiplied by a sign factor ($s=\pm 1$). In the latter case, one of the two non-zero elements is $\pm \frac{1}{2}(c_k + c_l)$ and the other one is $\pm \frac{1}{2}(c_k - c_l)$ [see the appendix for more details].

Clearly, existence of $K$ admissible linearly independent vectors $\vec{c}_k$, for which matrices $U_k = U(\vec{c}_k)$ are non-singular, together with condition (a) guarantees that if the original 2-observer $M \times N$ settings inequality $I(\vec{a}, \vec{b}) \leq 1$ is tight (is satisfied by $MN$ linearly independent vectors $a_s \otimes b_s, \ s=1,\ \ldots , \ MN$), then the extended inequality $I'(\vec{a}, \vec{b}, \vec{c}) = I(\vec{a}, U(\vec{c})\vec{b})\leq 1$ is also tight. Indeed, condition (a) guarantees that the extended inequality is satisfied by all admissible vertices $\vec{a} \otimes \vec{b} \otimes \vec{c}$. Moreover, existence of $K$ independent admissible vectors $\vec{c}_k$, for which matrices $U(\vec{c}_k)$ are non-singular, guarantees that there are $MNK$ linearly independent vectors $\vec{a}_s \otimes (U_k^{-1} \vec{b}_s) \otimes c_k$, which saturate the extended inequality. We can thus relax condition (b) without changing its implications. The relaxed condition requires that when observer $C$ considers $K$ measurement settings, $U(\vec{c})$ has to be non-singular for $K$ linearly independent admissible vectors $\vec{c}_k, \ k = 1, \ldots , \ K$ only.

We will see, however, that interesting extensions can be obtained via $U(\vec{c})$'s, which do not even satisfy the relaxed condition (b). In these cases, in order to check the tightness of the extended inequality, we will have to check the number of linearly independent vertices, which saturate it. If the number is $MNK$, then the generated inequality is tight, if not then the inequality is not tight.

\section{Examples}
\begin{enumerate}
\item Returning to our first example, we obtain the following $3$-observer inequality
\begin{equation}\label{a3}
\frac{1}{2}\left<((a_1 + a_2)b_1 + (a_1 - a_2)b_2)c_1 \right>  \leq 1
\end{equation}
This is the $2\times 2\times 2$ Bell inequality (A3) in Ref. \cite{ww}.
\item An example of a non-singular and non-diagonal matrices $U$ is
$$
U(\vec{c})=\left(\begin{array}{cc}
                  \frac12(c_1+c_2) & \frac12(c_1-c_2) \\
                  \frac12(c_1-c_2) & \frac12(c_1+c_2) \\
\end{array}\right),
$$
The matrix is invertible for all admissible vectors $\vec{c}$. Hence we get another tight $2\times 2\times 2$ Bell inequality
\begin{equation}\label{a4}
\frac12\left<a_1c_1(b_1+b_2)+a_2c_2(b_1-b_2)\right>\leq 1.
\end{equation}
This is inequality (A4) in Ref. \cite{ww}. One may notice that essentially the same $2\times 2\times 2$ inequality is generated from CHSH by a diagonal

$$
U(\vec{c})=\left(\begin{array}{cc}
                  c_1 & 0 \\
                  0 & c_2 \\
\end{array}\right),
$$

\item By changing the sign of one of the matrix elements in the previous example, we obtain
$$
U(\vec{c})=\left(\begin{array}{cc}
                  \frac12(c_1+c_2) & \frac12(c_1-c_2) \\
                  -\frac12(c_1-c_2) & \frac12(c_1+c_2) \\
\end{array}\right).
$$
This is again invertible for every admissible vector $\vec{c}$. Hence it gives another tight $2\times 2\times 2$ Bell inequality:
\begin{equation}\label{a5}
\frac12\left<a_1b_1c_2+a_1b_2c_1+a_2b_1c_1-a_2b_2c_2\right>\leq 1.
\end{equation}
It is inequality (A5) in Ref. \cite{ww}.

\item Consider
\begin{equation}
U(\vec{c})=\left(
\begin{array}{cc}
                  \frac12(-c_1+c_2) & \frac12(c_1+c_2) \\
                  c_1 & 0 \\
\end{array}
\right).
\end{equation}

Now, the linearly independent vertices $\vec{c}_1=(1,1)$,
and $\vec{c}_2=(1,-1)$ produce matrices
$$
U_1=\left(\begin{array}{cc}
           0 & 1\\
           -1 & 0\\
           \end{array}
           \right),\,
U_2=\left(\begin{array}{cc}
           -1 & 0\\
           -1 & 0\\
           \end{array}
           \right).
$$
Matrix $U_1$ is invertible but $U_2$ is not. Nevertheless, one can easily check that there are eight linearly independent admissible vectors, which satisfy the extended inequality. Thus the inequality is tight. It reads
\begin{equation}
\frac14\left<-3a_1b_1c_1+a_1b_1c_2+a_1b_2c_1+a_1b_2c_2+a_2b_1c_1+a_2b_1c_2+a_2b_2c_1+a_2b_2c_2\right>\leq
1,
\end{equation}
which can be recognized as inequality A2 in Ref. \cite{ww} (see also \cite{wbz}).

\item Finally, to complete the list of $3$-observer, $2$-settings per observer inequalities, we choose
\begin{equation}
U(\vec{c})=\left(
\begin{array}{cc}
                  c_1 & 0 \\
                  c_1 & 0 \\
\end{array}
\right).
\end{equation}
This time $U(\vec{c})$ is singular for all admissible vectors $\vec{c}$. Nevertheless, the obtained inequality, even if trivial, is tight. It is
$$
a_1b_1c_1 \leq 1
$$
which can be recognized as inequality (A1) in Ref. \cite{ww}.

\item Clearly, one can use more transformations successively, like in the example below. Here we put
\begin{eqnarray*}
b_1^{\prime}(\vec{c})&=&-\frac12[b_1(c_1+c_2)+b_2(c_1-c_2)],\\
b_2^{\prime}(\vec{c})&=&-\frac12[-b_1(c_1-c_2)+b_2(c_1+c_2)],\\
\end{eqnarray*}
and
\begin{eqnarray*}
a_1^{\prime}(\vec{d})&=&\frac12[a_1(d_1+d_2)+a_2(d_1-d_2)],\\
a_2^{\prime}(\vec{d})&=&\frac12[-a_1(d_1-d_2)+a_2(d_1+d_2)].\\
\end{eqnarray*}
This leads to the MABK inequality \cite{mabk}:
\begin{eqnarray}\label{four}\nonumber
&&a_1b_1c_1d_1-a_1b_1c_1d_2-a_1b_1c_2d_1-a_1b_2c_1d_1-a_2b_1c_1d_1-a_1b_1c_2d_2\\\nonumber
 &&
-a_1b_2c_1d_2-a_2b_1c_2d_2-a_1b_2c_2d_1-a_2b_1c_2d_1-a_2b_2c_1d_1+a_2b_2c_2d_2\\
 && +a_2b_2c_2d_1+a_2b_2c_1d_2+a_2b_1c_2d_2+a_1b_2c_2d_2\leq 4.
\end{eqnarray}
\end{enumerate}

Our extensions are not limited to the generation of previously known Bell inequalities.

\begin{ex}\rm Consider, e.g., the $4\times 4$ Bell inequalities found by
Gisin \cite{GISIN05}.
\begin{eqnarray}\label{4by4}\nonumber
(2 a_0 + a_1 + a2) b_0 + (a_0 - a_1 - a_2 - a_3) b_1 &  \\
+ (a_0 -a_1 - a_2 + a_3) b_2 + (-a_1 + a_2) b_3 & \leq 6,
\end{eqnarray}
and
\begin{eqnarray}\label{4by42}\nonumber
(-2a_0 + a_1 + a_2 + 2a_3)b_0 + (2a_0 + 2a_1 + a_2 + a_3)b_1 &
\\  + (a_0 - 2a_1 + 2a_2 + a_3)b_2 + (a_0 - a_1 - 2a_2 +
2a_3)b_3 & \leq 10.
\end{eqnarray}

\begin{enumerate}
\item
A simple transformation $b'_i (\vec{c})= c_i b_i$ is associated with $U(\vec{c})=\hbox{diag}(c_0,c_1,c_2,c_3)$. which is clearly non-singular for all admissible vectors $\vec{c}$. Thus, the resulting (rather trivial) extensions of the $4 \times 4$ inequalities are tight. They read
\begin{eqnarray}\label{4by4by4}\nonumber
(2 a_0 + a_1 + a2) b_0 c_0+ (a_0 - a_1 - a_2 - a_3) b_1 c_1&  \\
+ (a_0 -a_1 - a_2 + a_3) b_2 c_2 + (-a_1 + a_2) b_3 c_3 & \leq 6,
\end{eqnarray}
and
\begin{eqnarray}\label{4by4by42}\nonumber
(-2a_0 + a_1 + a_2 + 2a_3)b_0 c_0+ (2a_0 + 2a_1 + a_2 + a_3)b_1 c_1&
\\  + (a_0 - 2a_1 + 2a_2 + a_3)b_2 c_2 + (a_0 - a_1 - 2a_2 +
2a_3)b_3 c_3& \leq 10.
\end{eqnarray}

\item Using a more complex transformation given by
\begin{equation}
U(\vec{c})=\left(
\begin{array}{cccc}
                  \frac12(c_0+c_1) & \frac12(c_0-c_1) & 0 & 0 \\
                  \frac12(c_0-c_1) & -\frac12(c_0+c_1) & 0 & 0 \\
                  0 & 0 & \frac12(c_2+c_3) & \frac12(c_2-c_3) \\
                  0 & 0 & \frac12(c_2-c_3) & -\frac12(c_2+c_3) \\
\end{array}
\right).
\end{equation}

We also get tight $4\times 4\times 4$ Bell inequalities. This time inequalities read
\begin{eqnarray}\label{4by4by43}\nonumber
a_0(-b_0c_0-3b_0c_1-3b_1c_0+b_1c_1+b_2c_2+b_2c_3+b_3c_2-b_3c_3)&\nonumber  \\
+2a_1(b_0c_1-b_1c_0+b_2c_2-b_3c_3)
+2a_2(b_0c_0-b_1c_1+b_2c_3+b_3c_2)&\nonumber\\
-a_3(b_0c_0-b_0c_1-b_1c_0-b_1c_1-b_2c_2-b_2c_3-b_3c_2+b_3c_3) &\leq 12,
\end{eqnarray}
and
\begin{eqnarray}\label{4by4by44}\nonumber
2a_0(-2b_0c_1-2b_1c_0+b_2c_2-b_3c_3)+2a_2(b_0c_0-b_1c_1+2b_2c_3+2b_3c_2)&\\
+a_1(3b_0c_0-b_0c_1-b_1c_0-3b_1c_1-3b_2c_2-b_2c_3-b_3c_2+3b_3c_3)&\nonumber\\
+a_3(3b_0c_0+b_0c_1+b_1c_0-3b_1c_1+3b_2c_2-b_2c_3-b_3c_2-3b_3c_3)
&\leq 20.
\end{eqnarray}
\end{enumerate}
Inequalities
(\ref{4by4by4},\ref{4by4by42},\ref{4by4by43},\ref{4by4by44}) are new tight 3-observer Bell inequalities.

We have investigated the maximal violation of the $4\times 4$ (\ref{4by4},\ref{4by42}) and the $4\times 4\times 4$ inequalities (\ref{4by4by4},\ref{4by4by42},\ref{4by4by43},\ref{4by4by44}). The operators on the left-hand side of (\ref{4by4}) and (\ref{4by42}) attain their maximal mean values of 8.165 and 11.504, respectively, for the maximally entangled state of two qubits. The extension leading to (\ref{4by4by4}) and (\ref{4by4by42}) brings no gain in this respect, at least if one considers the mean values in the GHZ state, $\frac{1}{\sqrt{2}}(|000\rangle+|111\rangle)$. 

On the other hand, the maximum values of the left-hand sides of (\ref{4by4by43}) and (\ref{4by4by44}) attainable on a 3-particle GHZ state reach 23.008 in both cases. Due to the differently normalized right-hand sides here and in the original inequalities this value corresponds to 11.504 in inequalities (\ref{4by4by4}) and (\ref{4by4by42}). Thus by extending inequality (\ref{4by4}), we produced inequality (\ref{4by4by43}), which can be violated much stronger than the original (\ref{4by4}).

\end{ex}

Our example of a situation when quantum violation of the extended inequality can exceed the maximum possible quantum violation of the original inequality is not unique. A somewhat more elementary example of the same phenomenon emerges when one considers the CHSH inequality as a $2$-particle extension of a (trivial) $1$-particle inequality $I(\vec{a})=a_1 \leq 1$. The CHSH inequality can be expressed as $I(U(\vec{b}) \vec{a}) \leq 1$. For that, one can choose
\begin{equation}
U(\vec{b})=\left(
\begin{array}{cc}
                  \frac12(b_1+b_2) & \frac12(b_1-b_2)  \\
                  \frac12(b_1-b_2) & \frac12(b_1+b_2)
\end{array}
\right).
\end{equation}

\section{Conclusions}
We have described a method for generation of non-trivial extensions of Bell inequalities from $p$ to $p+1$ observers and formulated sufficient conditions for such extensions to produce tight inequalities from tight originals. Some of our extensions are interesting in the sense that correlations provided by entangled quantum states can violate the new inequalities more than the original ones. In fact, an extension can produce a non-trivial inequality from a trivial one (e.g. CHSH from inequality $a_1 \leq 1$). Moreover, our $3$-particle extensions of CHSH cover all the possible inequivalent $2 \times 2 \times 2$ correlation inequalities in dichotomic variables. This result may tempt one to conjecture that by our extensions, we can produce all inequivalent multi-observer correlation inequalities at least in case of dichotomic observables and two or (possibly) three settings per observer. At this stage we leave this as an open problem for future research.

The work is part of EU 6FP programmes QAP (no. IST-015848) and SCALA. M. \.{Z}ukowski
was supported by Wenner-Gren Foundations. M. Wie\'{s}niak is supported
by FNP stipends (START programme and within Professorial Subsidy
14/2003 for MZ), the National University of Singapore Grant No. R-144-000-206-112 and  the CQT grant no. R-710-000-010-271.

\section*{Appendix}
In this appendix we discuss the possible expressions of the elements of matrix $U(\vec{c})$.

When for some $(i,j)$, $u_{ij}(\vec{c}) \notin \{0, \pm 1 \}$ then with admissible $\vec{b}=(\ldots, b_j, \ldots)$ and $\vec{\textbf{b}}=(\ldots, -b_j, \ldots)$ at least one of the vectors $\vec{b}'(\vec{b}, \vec{c})$ and $\vec{b}'(\vec{\textbf{b}}, \vec{c})$ is not admissible. Moreover, when for a given admissible $\vec{c}$, the number of ones in a single row of $U(\vec{c})$ is not one, then there are admissible vectors $\vec{b}$, which are transformed into non-admissible vectors $\vec{b}'$. Thus, for each given admissible vector $\vec{c}$, the elements of $U(\vec{c})$ may only be $0 \hbox{ or } \pm 1$ and in each row of $U(\vec{c})$ there must be exactly one element equal to $1$.
This condition requires that by changing the sign of a single component of $\vec{c}$, one either leaves $u_{ij}(\vec{c})$ intact or changes it by $\pm 1$ or by $\pm 2$. With $u_{ij}(\vec{c})=\sum_k x^{ij}_k c_k$ this restricts the values of $x^{ij}_k$ to $0\hbox{ or } \pm 1 \hbox{ or } \pm \frac{1}{2}$. Moreover, we also need
$\sum_k|x^{ij}_k|=0\hbox{ or }1$. 

Consequently, if for some $k$, $x_k^{ij}=\pm 1$, then the other coefficients are zero; if $x_k^{ij}=\pm \frac12$ then one and only one of the other coefficients is $\pm \frac12$. In other words, if one of the elements of $U(\vec{c})$, $u_{ij}(\vec{c})=\sum_k x^{ij}_k c_k$ is not identically 0, then $u_{ij}(\vec{c})=\pm c_k \hbox { or }\pm\frac12(c_k\pm c_l)$. If $u_{ij}(\vec{c})=\pm c_k$ then the other elements in the same row are identically 0;  if $u_{ij}(\vec{c})=\pm \frac12(c_k\pm c_l)$ then we need another non-zero element in this row, namely $u_{ij'}(\vec{c})=\pm \frac12(c_k \mp c_l)$ for some $j' \neq j$. In this case, all the remaining elements in the row must be identically 0. Otherwise, we can choose an admissible vector such that there are two non-zero elements in this row.


\begin{thebibliography}{}
\bibitem{math}Mathematically, for a n-dimension polytope, the intersection of a
tangent hyperplane and the polytope is called face. Every
($n-1)$-dimensional face of the polytope is called a facet. In other words, a
facet is a hyperplane, which contains $n$ linearly independent
vertices of the polytope and such that the whole polytope is on its one side.
\end{thebibliography}
\end{document}